\documentclass[prb, twocolumn,amsmath,amssymb]{revtex4}
\usepackage{graphicx,subfigure,epsfig,verbatim,psfrag,amsmath,amssymb,color}
\usepackage{dcolumn}
\usepackage{bm}
\usepackage{epstopdf}
\usepackage{mathrsfs}
\usepackage{color}
\graphicspath {{figure/}}
\begin{document}
\title{Generalized Falicov-Kimball Models}
\author{Xiao-Hui Li\footnote{xlibb@connect.ust.hk} }
\author{Zewei Chen}
\author{Tai Kai Ng\footnote{phtai@ust.hk}}

\affiliation{Department of Physics, Hong Kong University of Science and Technology, Clear Water Bay, Hong Kong, China}
\date{\today}

\begin{abstract}
In this paper we extend the Falicov-Kimball model (FKM) to the case where the ``quasi-particles" entering the FKM are not ``ordinary" fermions. As an example we first discuss how the FKM can be generalized to the case with spin-dependent hopping. Afterward we discuss several cases where the ``quasi-particles" entering the FKM are Majorana fermions (extended Majorana-Falicov-Kimball Model (MFKM). Two examples of extended MFKM are discussed in detail: (i) a $p$-wave BCS superconductor on a bipartite lattice and (ii) a BCS-Anderson model. We also discuss the most general forms of extended MFKM, including a brief discussion on the case where the Majorana fermions represent spins, but not real fermion particles.
\end{abstract}
\maketitle
\section{Introduction}
 Falicov-Kimball Model(FKM) is one of the simplest exactly solvable interacting fermion models that displays a variety of rich phenomena such as metal insulator transition and charge/spin density wave order [\onlinecite{falicov1969simple,ramirez1970metal}]. In its simplest form, the model describes a system of itinerant (spin $\uparrow$) electrons interacting with localized (spin $\downarrow$) electrons [\onlinecite{falicov1969simple}]. The Hamiltonian is
\begin{equation} \label{Eq:HFKM}
 H = t\sum_{i,j} ( c_{i\uparrow}^\dagger c_{j\uparrow} + h.c.)+U\sum_{i} c_{i\uparrow}^\dagger c_{i\uparrow} c^\dagger_{i\downarrow} c_{i\downarrow} -\sum_{i\sigma}\mu_{\sigma} c^\dagger_{i\sigma} c_{i\sigma}
\end{equation}
 where electron hopping is allowed only for spin-$\uparrow$ electrons. $U$ denotes the on-site interaction between electrons and $\mu_{\sigma}$ is the chemical potential for spin-$\sigma$ electrons. The model is exactly solvable because the $\downarrow$-spin electrons are all localized and $n_{i\downarrow}$'s can be treated as $Z_2$ real numbers $(=0,1)$. The remaining Hamiltonian on the $\uparrow$-spin electrons is quadratic and exactly solvable. The model and its extensions have been studied in detail both analytically and numerically over the past decades  [\onlinecite{kennedy1994some,freericks2003exact,Haller2001,lemanski2002stripe,lebowitz1994long,zenker2010existence,vzonda2009phase}].

 It is interesting to note that several recent works have pointed out that FKM may be realized in cold atom systems as a tunable mass-imbalance Hubbard model, which has been realized experimentally with light and heavy atoms or with imbalanced left and right circular polarized lights or rotating gradient magnetic field  [\onlinecite{jaksch1999entanglement,liu2004spin,mandel2003coherent,yi2008state}].

 More recently, we showed that a class of $S=1/2$ lattice fermion models with equal spin pairing and on-site Hubbard interaction $U$ can be mapped into a Falicov-Kimball type model when the electron hopping $t$ equals the BCS pairing amplitude $\Delta$; the model has an FKM form when represented in terms of Majorana fermions [\onlinecite{chen2018exactly}]. As a unifying concept, we explore in this paper the general conditions for which a spin-$1/2$ fermion model with Hubbard interaction can be reduced to an FKM form. We consider the cases where the quasi-particles are (i) normal fermions and (ii) Majorana fermions. We note that in case (ii) the Majorana fermions may represent spins but not real fermions. As the term Majorana-Falicov-Kimball model was introduced in Ref. [\onlinecite{prosko2017simple}] in a different context, we name the model in case (ii) as the extended Majorana-FKM(MFKM) in the following text.  A general recipe for constructing the extended MFKM (or MFKM) is discussed. To illustrate, we study in detail the spin-dependent Haldane Hubbard model as an example in case (i). In case (ii) we first study a $p$-wave BCS superconductor with on-site Hubbard interaction $U$ [\onlinecite{chen2018exactly}] and then a two-band Anderson lattice type model with inter-band BCS pairing.

\section{Models}\label{Sec:Models}
In this section, we discuss how to construct generalized FKM with Hubbard-type on-site interactions. We first consider tight-binding models with spin-dependent hopping terms.

\subsection{Tight-binding Model with Spin-Dependent Hopping}\label{Sec:SpinU2}
 We consider a general tight-binding model with spin-dependent hopping and Hubbard on-site interaction,
\begin{equation}\label{Eq:Model_base}
\begin{split}
H =& \sum_{ i,j,\sigma,\sigma'} (t_{ij,\sigma\sigma'} c_{i,\sigma}^\dagger c_{j,\sigma'} +h.c.) + U \sum_{i} n_{i,\uparrow} n_{i,\downarrow} \\& - \mu \sum_{i} n_i
\end{split}
\end{equation}
where $\sigma = \uparrow,\downarrow$ represents different spin species, $t_{ij,\sigma\sigma'}$ is the matrix element for a spin-$\sigma$ fermion at site $i$ hopping to become a spin-$\sigma'$ fermion at site $j$, $U$ is the Hubbard on-site interaction and $\mu$ is the chemical potential.

 The hopping Hamiltonian can be written in a matrix notation as $H=\sum_{i,j}H_{ij}$, where $H_{ij} = \Psi_i^\dagger  \mathrm{T}_{ij} \Psi_j$, where $\Psi_i^\dagger = \{ c^\dagger_{i\uparrow},c_{i,\downarrow}^\dagger\}$ represents a fermion spinor and $\mathrm{T}_{ij}$ is $2\times 2$ matrix,
\begin{equation}
\mathrm{T}_{i,j} = \left(\begin{matrix}
       t_{ij,11} & t_{ij,12} \\[0.3em]
      t_{ij,21} & t_{ij,22}
     \end{matrix}\right), \ \
\end{equation}
with $t_{ij,12}=t_{ji,21}^*$ due to the Hermicity of the Hamiltonian. 

 We next introduce a site-dependent $\mathrm{SU(2)}$ rotation on the spinor,
\begin{equation}\label{Eq:SpinU2}
\tilde{\Psi}_i\equiv \left(\begin{matrix}
\tilde{c}_{i,\uparrow} \\
 \tilde{c}_{i,\downarrow}
\end{matrix}\right)= \mathcal{R}_{s,i} \left(\begin{matrix}
c_{i,\uparrow} \\
c_{i,\downarrow}
\end{matrix}\right), \
\mathcal{R}_{s,i}\equiv \left(\begin{matrix}
\alpha_i & \beta_i \\
-\beta^*_i & \alpha^*_i
\end{matrix}\right)
\end{equation}
where $\mathcal{R}_{s,i}$ are unitary matrices with $|\alpha_i|^2+|\beta_i|^2=1$. Correspondingly, $H_{ij}\rightarrow \tilde{\Psi}_i^\dagger  \tilde{\mathrm{T}}_{ij} \tilde{\Psi}_j$ where $\tilde{\mathrm{T}}_{i,j} = \mathcal{R}_{s,i} \mathrm{T}_{ij}\mathcal{R}_{s,j}^\dagger $ in the rotated basis. As the Hubbard interaction term is spin-rotational invariant, the interaction term $U \tilde{n}_{\uparrow} \tilde{n}_{\downarrow}=Un_{\uparrow}n_{\downarrow}$ remains unchanged in the spin-rotated basis. The model becomes a FKM if we can find a set of $\mathcal{R}_{s,i}$'s such that
\begin{equation}\label{Eq:ExactCondition}
\tilde{\mathrm{T}}_{ij} = \mathcal{R}_{s,i} \mathrm{T}_{ij} \mathcal{R}_{s,j}^\dagger =\left(\begin{matrix}
\tilde{t}_{i,j} & 0 \\
0 & 0
\end{matrix}\right),
\end{equation}
i.e., the hopping Hamiltonian $H_{ij} = \Psi_{i}^\dagger \mathrm{T}_{ij} \Psi_{j} =\tilde{\Psi}_{i}^\dagger \tilde{\mathrm{T}}_{ij} \tilde{\Psi}_{j}$ contains only one hopping specie in the rotated basis $\tilde{\Psi}$ which is precisely the FKM. To illustrate, we discuss two examples in the following:


\subparagraph{Case a}Consider the simplest case where the system has translational symmetry and the rotation matrix $\mathcal{R}_{s,i}$ is site-independent. In this case it is easy to see that any hopping Hamiltonian $\mathrm{T}_{i,j}$ ($(i,j)=$ nearest neighbors) of form
\begin{equation}\label{Eq:Generalhij}
\mathrm{T}_{i,j}= \tilde{t}\left(\begin{matrix}
       |\alpha|^2 & \alpha^*\beta\\[0.3em]
      \alpha\beta^* & |\beta|^2
     \end{matrix}\right)
\end{equation}
 is exactly solvable. A simple example is
\begin{equation}
     \mathrm{T}_{i,j} = \tilde{t}\left(\begin{matrix}
       \frac{1}{2} & \frac{e^{i\theta}}{2}\\[0.3em]
      \frac{e^{-i\theta}}{2} & \frac{1}{2}
     \end{matrix}\right), \
     \mathcal{R}_{s,i}\equiv {1\over\sqrt{2}}\left(\begin{matrix}
     1 & e^{i\theta} \\
-e^{-i\theta} & 1
\end{matrix}\right)
\end{equation}
corresponding to $\alpha=1/\sqrt{2}, \beta=e^{i\theta}/\sqrt{2}$.

\subparagraph{ Case b.} We consider a bipartite lattice where $H_{i,j}$ describes hopping between two nearest neighbor sites belonging to different sublattices $A$ and $B$. In this case a natural possibility is that the rotational matrix $\mathcal{R}_{s,i}$ for the sites $i$ on two sublattices are different, i.e. $\mathcal{R}_{s,A} \neq \mathcal{R}_{s,B}$. 
In this case a sufficient condition that the system can be reduced to a FKM-type Hamiltonian is
\begin{equation}
\begin{split}
&\tilde{\mathrm{T}}_{i\in A,j \in B} = \mathcal{R}_{s,A} \mathrm{T}_{i,j} \mathcal{R}_{s,B}^\dagger =\left( \begin{matrix}
       \tilde{t}_{A, B} & 0 \\[0.3em]
      0 & 0
     \end{matrix} \right), \\
&\tilde{\mathrm{T}}_{i\in B,j \in A} = \mathcal{R}_{s,B} \mathrm{T}_{i,j} \mathcal{R}_{s,A}^\dagger = \left( \begin{matrix}
       \tilde{t}_{B, A} & 0 \\[0.3em]
      0 & 0
     \end{matrix}\right),
\end{split}
\end{equation}
and $\mathrm{T}_{i,j}$'s must be of form
\begin{equation}
\begin{split}
\mathrm{T}_{i\in A,j \in B} = \tilde{t}_{A,B} \left(\begin{matrix}
       \alpha_{A}^*\alpha_B & \alpha_A^*\beta_B \\[0.3em]
      \alpha_A\beta_B^* & \beta_A\beta_B^*
     \end{matrix}\right), \\
\mathrm{T}_{i\in B,j \in A} = \tilde{t}_{B,A} \left(\begin{matrix}
       \alpha_{B}^*\alpha_A & \alpha_B^*\beta_A \\[0.3em]
      \alpha_B\beta_A^* & \beta_B\beta_A^*
     \end{matrix}\right).
\end{split}
\end{equation}

 Spin-dependent hopping can be realized in systems with spin-orbit coupling, although it is not easy to realize the FKM limit where one of the fermion species band becomes flat. The quasi-particles are (site-dependent) spin-flipped fermions in these models. In Sec. \ref{Sec:Examples} we shall study an example of spin-dependent hopping models, and propose a simple mean-field theory that can extrapolate the system away from the FKM limit.

\subsection{Extended Majorana Falicov-Kimball Model}\label{Sec:MaFKM}
In the previous subsection, we discussed general forms of spin-dependent hopping Hubbard model which can be mapped to FKM, in which the quasi-particles are fermions. In this subsection, we explore the situation where the quasi-particles are Majorana fermions.

We introduce spin-$\sigma$ Majorana fermions $\gamma_{\sigma}, \eta_{\sigma}$ related to ordinary fermions $c_{\sigma}, c^\dagger_{\sigma}$ by
\begin{subequations}\label{Eq:DefMaj}
\begin{align}
&\gamma_{\sigma} = \frac{1}{2}(c_{\sigma}+c_{\sigma}^\dagger) \\
&\eta_{\sigma} = \frac{i}{2} (c_{\sigma}-c_{\sigma}^\dagger)
\end{align}
\end{subequations}
The Majorana fermions are self-conjugate with the properties $\gamma_\sigma^2 =\eta_\sigma^2 = 1/4$.

 We can now write down an FKM with Majoranas as quasi-particles  [\onlinecite{majorana1937teoria}]. The extended Majorana-Falicov-Kimball model (MFKM) Hamiltonian is
 \begin{equation}\label{Eq:MaFKM}
H = -i \sum_{i,j \sigma,\sigma'} g_{ij,\sigma\sigma'} \gamma_{i,\sigma} \gamma_{j,\sigma'} + U\sum_{i}(2i\gamma_{i\uparrow}\eta_{i\uparrow} 2i \gamma_{i\downarrow} \eta_{i\downarrow} ).
\end{equation}
Similar to usual FKM, the kinetic part of this Hamiltonian contains only one of the Majorana species $\gamma_{i,\sigma}$'s. Therefore, $[\eta_{l\uparrow}\eta_{l\downarrow},H]=0$ and can be replaced by $C$-numbers. As $(4i\eta_{\uparrow}\eta_{\downarrow})^2 \equiv 1$, $2i\eta_{\uparrow} \eta_{\downarrow} = \pm \frac{1}{2}$ is a $Z_2$ number. The remaining Hamiltonian reduces to a quadratic form for the $\gamma_{i,\sigma}$ Majorana fermions and is thus exactly solvable. Notice that $4U i\gamma_\uparrow \eta_\uparrow i \gamma_\downarrow \eta_\downarrow=U(n_{i,\uparrow}-\frac{1}{2})(n_{i,\downarrow}-\frac{1}{2})$ represents Hubbard interaction in original fermion representation. Notice that the usual fermion chemical potential term $\mu\sum_{i,\sigma}c^\dagger_{i\sigma}c_{i\sigma}=2i\mu\sum_{i,\sigma}\gamma_{i\sigma}\eta_{i\sigma}$ introduces hybridization between $\gamma$ and $\eta$ fermions and is set to zero in the extended MFKM, i.e., the extended MFKM describes a particle-hole symmetric system with on-site Hubbard interaction.

To find the physical fermion models that correspond to the extended MFKM, we first write down a general BCS-Hubbard model in the fermion basis,
\begin{equation}\label{Eq:Model}
\begin{split}
H =& \sum_{ i,j,\sigma,\sigma'} (t_{ij,\sigma\sigma'} c_{i,\sigma}^\dagger c_{j,\sigma'} +h.c.) +(\Delta_{ij,\sigma,\sigma'} c_{i,\sigma}^\dagger c_{j,\sigma'}^\dagger +h.c.)\\&+ U \sum_{i} (n_{i,\uparrow} - \frac{1}{2}) (n_{i,\downarrow} -\frac{1}{2})
\end{split}
\end{equation}
where $\sigma = \uparrow,\downarrow$ represents different spin species, $t_{ij,\sigma\sigma'}$ is the spin-dependent hopping matrix from sites $i$ to site $j$ and $\Delta_{ij,\sigma\sigma'}$ is the spin-dependent pairing potential between particles in sites $i,j$ and $U$ is the Hubbard on-site interaction. With the Majorana fermions introduced in Eq. (\ref{Eq:DefMaj}), the Hamiltonian (\ref{Eq:Model}) can be rewritten as,
\begin{equation}\label{Eq:HMajorana}
\begin{split}
H=2i\sum_{\langle i,j\rangle,\sigma\sigma'} &[\mathrm{Im} (t_{ij,\sigma\sigma'}+\Delta_{ij,\sigma\sigma'}) \gamma_{i,\sigma}\gamma_{j,\sigma'} \\ &+\mathrm{Im}(t_{ij,\sigma\sigma'}-\Delta_{ij,\sigma\sigma'})\eta_{i,\sigma} \eta_{j,\sigma'}] \\+2i\sum_{\langle i,j \rangle,\sigma\sigma'}&[\mathrm{Re}(t_{ij,\sigma\sigma'}-\Delta_{ij,\sigma\sigma'})\gamma_{i,\sigma}\eta_{j,\sigma'} \\ -&\mathrm{Re}(t_{ij,\sigma\sigma'}+\Delta_{ij,\sigma\sigma'})\eta_{i,\sigma} \gamma_{j,\sigma'}]\\+ U\sum_{i}(2i&\gamma_{i,\uparrow}\eta_{i,\uparrow})(2i\gamma_{i,\downarrow}\eta_{i,\downarrow}).
\end{split}
\end{equation}
  In particular, the model reduces to the extended MFKM described by Eq. (\ref{Eq:MaFKM}) when only two of the four Majorana fermions appear in the kinetic term that connects different sites, i.e. when $\mathrm{Re}(t_{ij,\sigma\sigma'})=\mathrm{Re}(\Delta_{ij,\sigma\sigma'})=0$ and $\mathrm{Im}(t_{ij,\sigma\sigma'})=\pm \mathrm{Im}(\Delta_{ij,\sigma\sigma'})$.

  Other forms of extended MFKM can be constructed through a unitary transformation as in the case of spin-dependent hopping models. In the following, we shall explore the most general form of BCS-Hubbard model that can be mapped to extended MFKM.

  To start with, we first consider the particle-hole transformation, which can be viewed as an $\mathrm{SU(2)}$ rotation in the pseudospin space (or a local Bogoliubov transformation) given by
\begin{equation}\label{Eq:PHU2}
\left( \begin{matrix}
\tilde{c}_{\uparrow} \\
\tilde{c}_{\downarrow}^\dagger
\end{matrix}\right) =  \mathcal{R}_{ph} \left( \begin{matrix}
c_\uparrow \\
c_\downarrow^\dagger
\end{matrix}\right),\ \mathcal{R}_{ph}\equiv \left(\begin{matrix}
a & b \\
-b^* & a^*
\end{matrix} \right)
\end{equation}
with $|a|^2+|b|^2=1$. Notice that $s_z=n_{\uparrow}-n_{\downarrow}$ is invariant under this transformation. Using the fact that $(n_{\uparrow}-{1\over2})(n_{\downarrow}-{1\over2})={1\over4}-{1\over2}(s_z)^2$, it is easy to see that the Hubbard interaction is invariant under $\mathcal{R}_{ph}$, i.e.
\[
U(2i\gamma_{i,\uparrow}\eta_{i,\uparrow})(2i\gamma_{i,\downarrow}\eta_{i,\downarrow})=
U(2i\tilde{\gamma}_{i,\uparrow}\tilde{\eta}_{i,\uparrow})(2i\tilde{\gamma}_{i,\downarrow}\tilde{\eta}_{i,\downarrow}). \]
where $\gamma(\tilde{\gamma})$ and $\eta(\tilde{\eta})$ are the Majorana fermions constructed from the $c (\tilde{c})$ fermions, respectively.
Notice however that the charge carried by the $\tilde{c}$- fermions are in general not equal to the charge carried by the original $c$- fermions under $\mathcal{R}_{ph}$ (see further discussion in Sec. \ref{Sec:AndersonLattice}). This is different from the spin-dependent FKM discussed in the previous subsection, where the charge carried by fermions remain unchanged under spin rotation.

 In the Majorana basis $\Gamma_i^T = \{\gamma_{i,\uparrow},\gamma_{i,\downarrow},\eta_{i,\uparrow},\eta_{i,\downarrow} \}$ , the particle-hole rotation described by $\mathcal{R}_{ph}$ can now be represented by a $4\times 4$ real matrix $\mathcal{P}_{ph}$, with
\begin{equation}\label{Eq:P_ph}
\tilde{\Gamma} = \mathcal{P}_{ph}\Gamma,\  \mathcal{P}_{ph} = \theta_0 \rho^0 \sigma^0 + i \theta_y \rho^y \sigma^0+ i \theta_x \rho^x \sigma^y + i\theta_z \rho^z \sigma^y
\end{equation}
where $\rho^{(x,y,z)}$ and $\sigma^{(x,y,z)}$ are the Pauli matrices acting on the charge-$\{\gamma,\eta\}$ and spin-$\{\uparrow, \downarrow\}$ subspaces, respectively. The parameters $\{\theta_{i}\} (i=0,x,y,z)$ corresponding to Eq. (\ref{Eq:PHU2}) are $\{\theta_0 = \mathrm{Re}(a),\ \theta_y = \mathrm{Im}(a),\  \theta_x = \mathrm{Im}(b),\  \theta_z = \mathrm{Re}(b)\ \}$. Under this basis,  one can rewrite the hopping Hamiltonion as $H_{ij} = \Gamma_i^T \mathrm{T}_{i,j} \Gamma_j$ where  $\mathrm(T)_{i,j}=-\mathrm(T^T)_{j,i}$. 

Besides of the $\mathrm{SU(2)}$ rotation in pseudospin space, Hubbard interaction term is also obviously invariant under spin-rotation. And in Majorana basis $\Gamma$, spin-rotation described by Eq. (\ref{Eq:SpinU2}) is represented as
\begin{equation}\label{Eq:P_s}
\tilde{\Gamma} = \mathcal{P}_s \Gamma,\  \mathcal{P}_s= \phi_0 \rho^0\sigma^0 + i\phi_x \rho^y \sigma^x + i\phi_y \rho^0\sigma^y + i\phi_z\rho^y\sigma^z
\end{equation}
where $\{ \phi_0 = \mathrm{Re}(\alpha),\ \phi_x =\mathrm{Im}(\beta),\  \phi_y = \mathrm{Re}(\beta) ,\ \phi_z =\mathrm{Im}(\alpha)\ \}$, corresponding to the parametrization in Eq. (\ref{Eq:SpinU2}). 

The spin-rotation and particle-hole transformations above can be combined into a general $\mathrm{SO(4)}$ transformation as $\mathrm{SO(4)}\sim \mathrm{SU(2)}\otimes \mathrm{SU(2)}$. Therefore, the Hubbard interaction term is invariant under such $\mathrm{SO(4)}$ rotation described by $\tilde{\Gamma} = \mathcal{P}\Gamma$. Then one should have $\mathcal{P}=\mathcal{P}_{ph}\otimes \mathcal{P}_{s}$ [\onlinecite{xu2010majorana}].

 Following previous discussions, for any BCS-Hubbard Hamiltonian of form (\ref{Eq:Model}), if it can be transformed to a Hamiltonian of form (\ref{Eq:MaFKM}) through an $\mathrm{SO(4)}$ rotation on the basis $\Gamma$ , then the Hamiltonian is exactly solvable.
 In the Majorana representation, we look for transformations of the Majorana basis $\tilde{\Gamma}_i = \mathcal{P}_i\Gamma_i$ and $\tilde{\mathrm{T}}_{i,j} = \mathcal{P}_i \mathrm{T}_{i,j} \mathcal{P}_j^T$ such that
\begin{equation}\label{Eq:ExactMaFKM}
\tilde{\mathrm{T}}_{i,j} = \left(\begin{matrix} \tilde{t}_{11ij} & \tilde{t}_{12ij} & 0 &0  \\[0.3em]
    \tilde{t}_{21ij} & \tilde{t}_{22ij} & 0 & 0\\[0.3em]
    0 & 0 &0 & 0 \\[0.3em]
    0 & 0 & 0 &0 \\[0.3em]
    \end{matrix}\right).
\end{equation} 
We note that as in the cases discussed in Sec. \ref{Sec:SpinU2} , the rotational matrix $\mathcal{P}_i$ can be site dependent. We now consider two simple examples of extended MFKM in the following. More examples will be discussed in Sec. \ref{Sec:Examples}.

\subparagraph{Case a.} We consider the simplest case where the system has translational symmetry and the rotation matrix $\mathcal{P}=\mathbb{I}$ is the same for all sites. This corresponds to the case where $t_{\sigma\sigma'}$ and $\Delta_{\sigma\sigma'}$ are pure imaginary numbers with $t_{\sigma\sigma'}=\pm\Delta_{\sigma\sigma'}$ in Eq.\ (\ref{Eq:HMajorana}).

\subparagraph{Case b.} Consider a bipartite lattice system in which the rotation matrix $\mathcal{P}$ differs at sublattice A and sublattice B, then a sufficient condition for the $\mathrm{T}_{i,j}$ is
\begin{equation}
\begin{split}
&\mathrm{T}_{A\to B } = \mathcal{P}_A^T \tilde{\mathrm{T}}_{i,j} \mathcal{P}_B \\
& \mathrm{T}_{B\to A } = \mathcal{P}_B^T \tilde{\mathrm{T}}_{j,i} \mathcal{P}_A
    \end{split}
\end{equation}
where $\tilde{\mathrm{T}}_{i,j}$ has the form of Eq. (\ref{Eq:ExactMaFKM}).

As an example, we take $\mathcal{P}_A =\mathbb{I}$ for sublattice A while $\mathcal{P}_B = \mathcal{P}_s\otimes\mathcal{P}_{ph} = \mathbb{I}\otimes i\rho^y\sigma^0$ for sublattice B. Therefore, the Majorana fermion basis remains the same, i.e. $\tilde{\Gamma}_A = \Gamma_A$ at sublattice A; however the basis is transformed as $\{\tilde{\gamma}_{\uparrow},\tilde{\gamma}_{\downarrow},\tilde{\eta}_{\uparrow},\tilde{\eta}_{\downarrow}\} = \{-\eta_\uparrow,-\eta_\downarrow,\gamma_{\uparrow},\gamma_\downarrow \}$ at sublattice B. The corresponding hopping matrix $\mathrm{T}_{i,j}$ in the original basis has the form
\begin{equation}
\begin{split}
\mathrm{T}_{A\to B} =  &\left(\begin{matrix} 0 & 0 & 0 & \tilde{t} \\[0.3em]
    0 & 0&-\tilde{t} & 0\\[0.3em]
    0 & 0 &0 & 0 \\[0.3em]
    0 & 0 & 0 &0 \\[0.3em]
    \end{matrix}\right), \\
    \mathrm{T}_{B\to A} =  &\left(\begin{matrix} 0 & 0 & 0 & 0 \\[0.3em]
    0 & 0 & 0 & 0\\[0.3em]
    0 & -\tilde{t} & 0 & 0 \\[0.3em]
    \tilde{t} & 0 & 0 &0 \\[0.3em]
    \end{matrix}\right),
 \end{split}
\end{equation}
   corresponding to the case where $t_{\sigma\sigma'}$ and $\Delta_{\sigma\sigma'}$ are real numbers and $t_{ij,\sigma\sigma'}=-\Delta_{ij,\sigma\sigma'} $ for $i (j)\in A (B)$ and  $t_{ij,\sigma\sigma'}=\Delta_{ij,\sigma\sigma'}$ for $i (j) \in B(A)$ in Hamiltonian(\ref{Eq:HMajorana}). This corresponds to the model studied in Ref.  [\onlinecite{chen2018exactly}].

\subsection{BCS - Anderson Model}\label{Sec:AndersonModel}
The exact solvability of models described in the above subsections relies on the fact that the quadratic part of the Hamiltonian generates two spin-split bands, one of which is flat and the Hubbard interacting term is invariant under spin rotation and particle-hole transformation in the space spanned by the two bands.  When multiple band systems are considered, the Hubbard interacting term will not be invariant under inter-band mixing in general. However, there is a special case of two-band models where the Hubbard interaction acts only on electrons in one band--the Anderson lattice model [\onlinecite{anderson1961localized}]. We consider here the BCS - Anderson lattice model with Hamiltonian
\begin{equation}\label{Eq:HAnderson}
\begin{split}
H = &\sum_{\langle i,j \rangle,\tau} (t_{\tau} c^\dagger_{i,\tau} c_{j,\tau}+ h.c.)-\sum_{i,\tau} \mu_{\tau} (c^\dagger_{i,\tau}c_{i,\tau})\\&+ \sum_{i,\tau,\sigma} (V_{1,\tau,\sigma} c^\dagger_{i,\tau} f_{i,\sigma}+V_{2,\tau,\sigma} c^\dagger_{i,\tau} f_{i,\sigma}^\dagger +h.c.)\\& + U\sum_{i} (f_{i,\uparrow}^\dagger f_{i,\uparrow}-\frac{1}{2})(f_{i,\downarrow}^\dagger f_{i,\downarrow}-\frac{1}{2})
\end{split}
\end{equation}
where $c(c^\dagger)$ is the annihilation (creation) operator for the mobile $c$-fermions and $f(f^\dagger)$ is the corresponding operator for the (immobile) $f$-fermions. $\tau$ denotes the spin for $c$-fermions, while $\sigma$ denotes the spin for $f$-fermions. To make the model exactly solvable, we have introduced an inter-band BCS pairing term $V_2$ into $H$. Notice that a spin-dependent chemical potential $\mu_{\tau}$ is also added to the $c$-fermions but the on-site energy is set to be zero for the $f$-fermions.

In analog to the discussions in Sec. \ref{Sec:MaFKM}, we can write down the hybridization matrix between the conducting and the $f$- fermions in the Majorana basis,
\begin{equation}\label{Eq:Majorana}
\begin{split}
\gamma_{i,\tau} = &\frac{1}{2}(c_{i,\tau} +c^\dagger_{i,\tau}),\\
\eta_{i,\tau} = &\frac{i}{2} (c_{i,\tau} -c^\dagger_{i,\tau}),\\
\chi_{i,\sigma} = &\frac{1}{2}(f_{i,\sigma} +f^\dagger_{i,\sigma}),\\
\xi_{i,\sigma} = &\frac{i}{2}(f_{i,\sigma} -f^\dagger_{i,\sigma}).
\end{split}
\end{equation}
In the $\Gamma_{i}=\{\gamma_{i,\uparrow},\gamma_{i,\downarrow},\eta_{i,\uparrow},\eta_{i,\downarrow}\}$ and $\Pi_i = \{\chi_{i,\uparrow},\chi_{i,\downarrow},\xi_{i,\uparrow},\xi_{i,\downarrow}\}$ bases, the coupling between the $c$- electron and $f$-fermions (i.e. the second line in Hamiltonian (\ref{Eq:HAnderson})) can be written as $H_c = \Gamma_{i} h_{c,i} \Pi_i$ and the model becomes exactly solvable if only two of the four Majorana fermions appear in the coupling matrix $h_{c,i}$. More generally we may have $h_{c,i}\rightarrow \tilde{h}_{c,i} = \mathcal{P}_{c}^T h_{c,i} \mathcal{P}_f$ where $\mathcal{P}_{c}$ and $\mathcal{P}_f$ are the rotation matrices as defined in previous sections acting on the $c$- and $f$- electrons, respectively. The general form of $\tilde{h}_{c,i}$ where the model becomes exactly solvable is, therefore 
\begin{equation}
\label{bcsa}
\tilde{h}_{c,i} = \mathcal{P}_c^Th_{c,i} \mathcal{P}_f = \left(\begin{matrix}
\tilde{t}_{11} & \tilde{t}_{12} & 0 &0\\
\tilde{t}_{21} & \tilde{t}_{22} & 0 &0\\
0 & 0 & 0 &0\\
0 & 0 & 0 &0
\end{matrix}\right).
\end{equation}
 The Hubbard interaction term in Eq. (\ref{Eq:HAnderson}) can be rewritten as $U\sum_{i}(2i\tilde{\chi}_{i,\uparrow}\tilde{\chi}_{i,\downarrow})(2i\tilde{\xi}_{i,\uparrow} \tilde{\xi}_{i,\downarrow})$, where $(2i\tilde{\xi}_{i,\uparrow} \tilde{\xi}_{i,\downarrow}) = D =\pm \frac{1}{2}$ is a good quantum number when Eq.\ (\ref{bcsa}) is satisfied. The model becomes exactly solvable.

\subsection{Kitaev Spin Model}\label{Sec:Kitaev model}
The extended MFKM may represent some exactly solvable spin models since Majorana fermions can be used to represent spins, not only canonical fermions. For spin $S=1/2$, the spin operators can be represented as
 \[
    S_x=ia_xb; \ S_y=ia_yb; \ S_z=ia_zb; \]
    with constraint $a_xa_ya_zb=1$. It has been shown by Chen and Nussinov [\onlinecite{chen2008exact}] and Feng, Zhang, and Xiang [\onlinecite{feng2007topological}] that the Kitaev spin-$1/2$ model on Honeycomb lattice [\onlinecite{kitaev2006anyons}] can be mapped to an extended  MFKM with the help of the Jordan-Wigner transformation. The details of the mapping can be found in references [\onlinecite{chen2008exact, feng2007topological}] and we shall not reproduce the calculation here. It is interesting to note that a 3-dimensional version of the Kitaev model has been constructed by Miao {\em et al.} [\onlinecite{miao2018exact}] with the help of the extended MFKM representation.

\section{Examples of FKM and extended MFKM}\label{Sec:Examples}
   To illustrate how the theory discussed in Sec. \ref{Sec:Models} can be applied in realistic models,  we study in this section several examples of FKM and extended MFKM. We first consider the Falicov-Kimball-Haldane model which is an extension of FKM with spin-dependent hopping. Using this example, we shall illustrate how a mean-field theory can be constructed to compute the properties of the system beyond the FKM (exactly solvable) limit.

\subsection{Falicov-Kimball Haldane Model }\label{Sec:FKMHaldane}
We first consider a spin-dependent Haldane Hubbard model, which at FKM limit is exactly solvable. Our study is motivated by the interests in realizing the topological Haldane model on cold atom system and that FKM may be realized in cold atom systems as a tunable mass-imbalance Hubbard model [\onlinecite{jotzu2014experimental, jotzu2015creating}]. It would be interesting if one can  realize the Haldane-Hubbard model in the FKM limit. We consider the Hamiltonian
\begin{equation}\label{HU}
\begin{split}
H = &-\sum_{\langle i,j \rangle,\sigma} (t_{1,\sigma\sigma'} c_{i,\sigma}^\dagger c_{j,\sigma'}+h.c.)+ \sum_i  M_i n_i -\mu \sum_i n_i\\& -\sum_{\langle\langle i,j \rangle\rangle,\sigma} (t_{2,\sigma\sigma'} e^{i\phi_{i,j}} c_{i,\sigma}^\dagger c_{j,\sigma'}+h.c.) \\&+U\sum_{i} (n_{i,\uparrow}-\frac{1}{2}) (n_{i,\downarrow}-\frac{1}{2}) 
\end{split}
\end{equation}
where $\sigma=\uparrow, \downarrow$, $t_{1,\sigma\sigma'}$ and $t_{2,\sigma\sigma'}$ describe spin-dependent hopping between nearest neighbor sites and next nearest neighbor sites for the $c$-fermions, respectively. $\phi_{i,j} = \phi (-\phi)$ for next nearest neighbor hopping between $A (B)$ sub-lattice sites. $M_i=M(-M)$ is a staggered on-site potential energy for $A(B)$ sub-lattice sites $i$. In the following calculation, we fix the staggered potential $M$ and chemical potential $\mu$ to be zero (i.e. half-filled bands) for simplicity.

When the hopping Hamiltonian satisfies the condition  given by Eq. (\ref{Eq:ExactCondition}), the Hamiltonian can be transformed to a standard FKM which can be exactly solved since $(\tilde{n}_{\downarrow}-\frac{1}{2})$ is a good quantum number and can be represented by a $Z_2$ number $D_i=\pm1/2$. We assume here that the spin-$\uparrow$ is the itinerant specie while the spin-$\downarrow$ fermions are localized. From the results of previous theoretical studies  [\onlinecite{watson1995ground, freericks2003exact,kennedy1994some}], we know that at half fillings the bipartite system favors staggered configuration  $D_i=+(-)1/2$ for $A(B)$ sublattices.  This generates an effective on-site potential difference $M_{eff}=U/2$ for fermions on $A$ and $B$ sublattices respectively, which can drive a topological phase transition in Haldane model [\onlinecite{haldane1988model}]. The topological region is given by $-3\sqrt{3}t_{2,\uparrow}\sin{\phi}< M_{eff} < 3\sqrt{3}t_{2,\uparrow}\sin{\phi}$.

\begin{figure}[h]
\centering
\mbox{
\subfigure{
\includegraphics[width=0.5\columnwidth, height=1.5in]{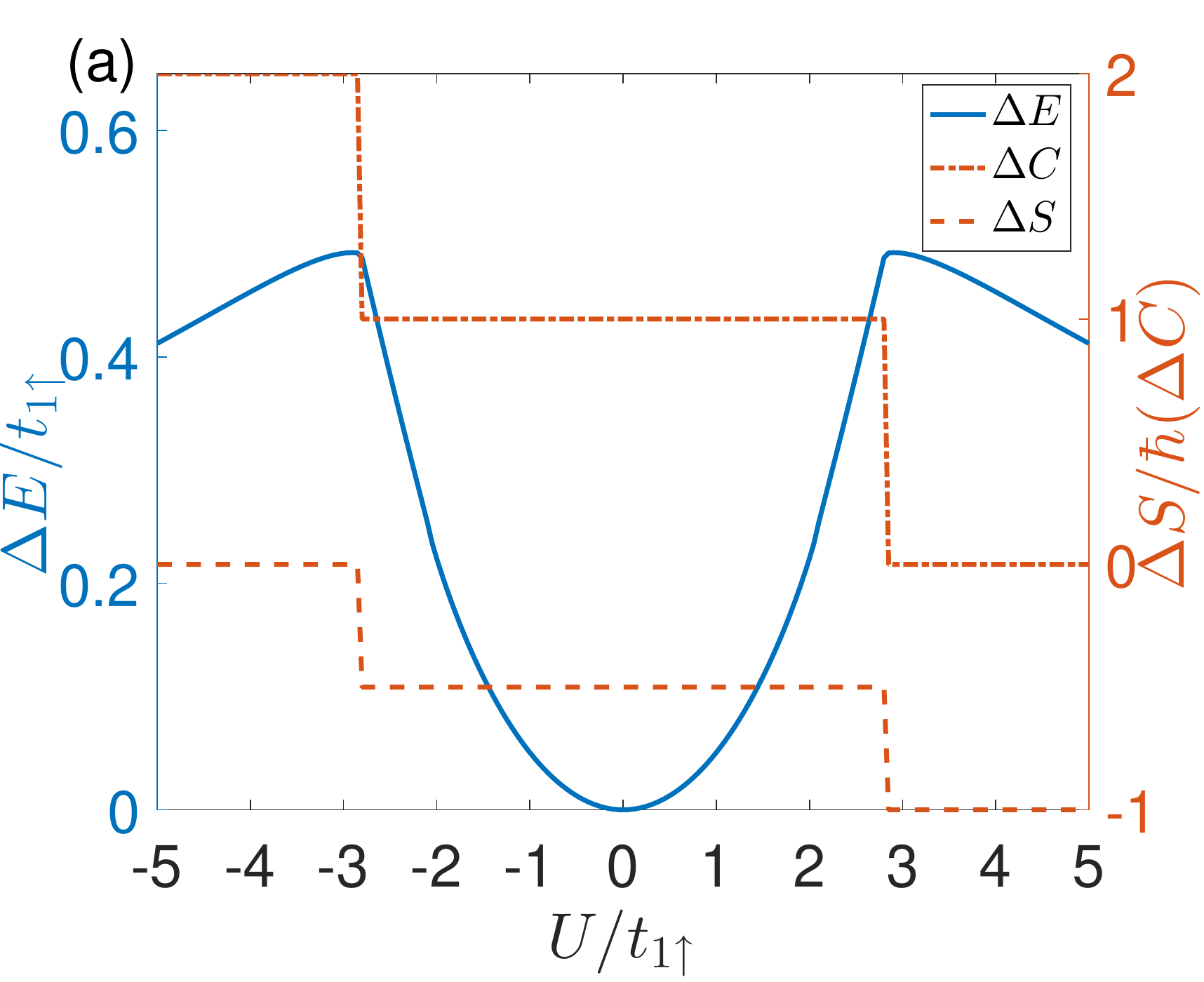}}
\subfigure{
\includegraphics[width=0.5\columnwidth, height=1.5in]{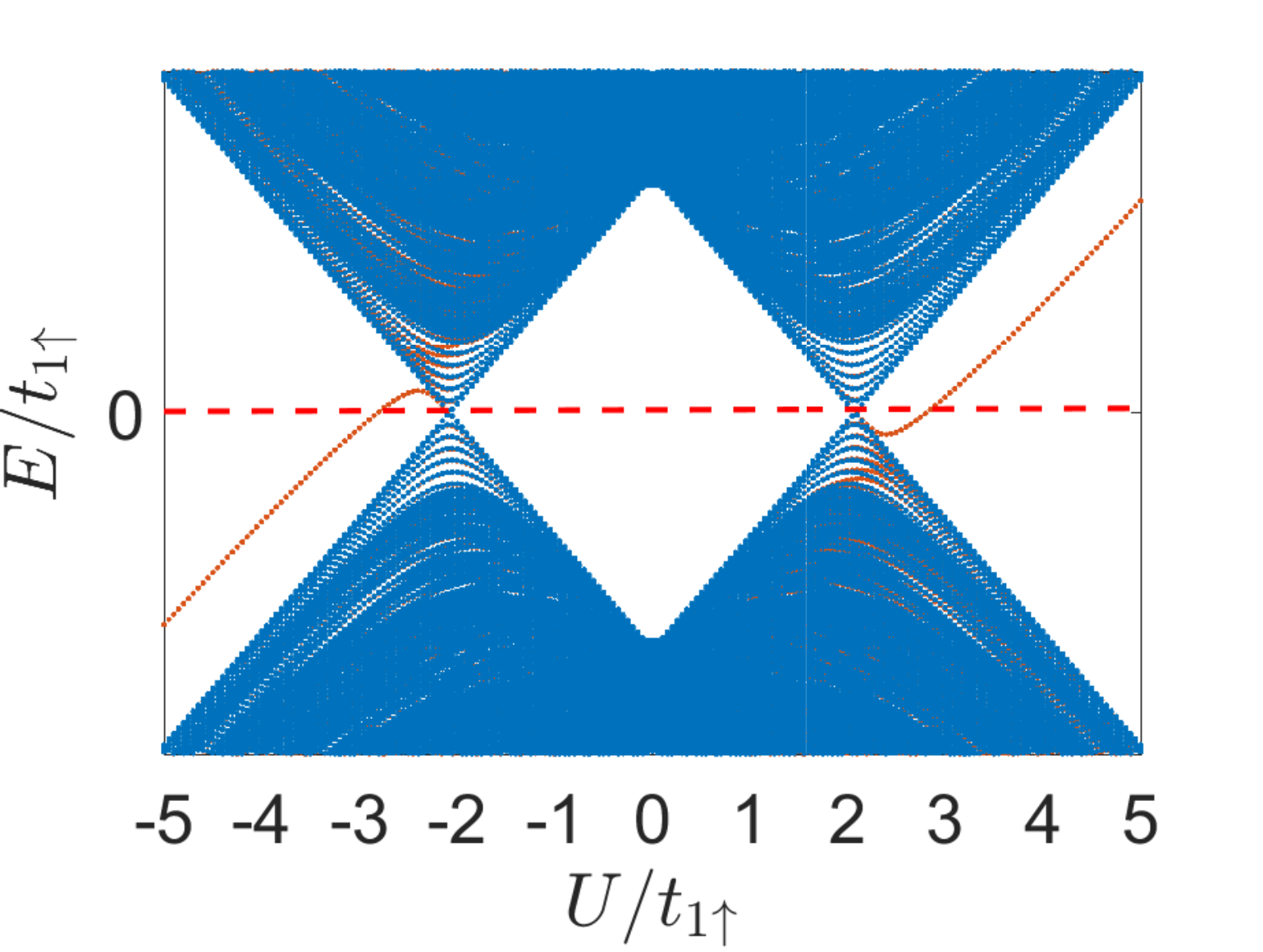}}
}
\caption{ (a) Solitonic excitation properties. The solid curve corresponding to label on the left is the solitonic excitation energy $\Delta E/t_{1\uparrow}$. The dashed curve corresponding to the label on the right is the spin carried by the excitation $\Delta S/ \hbar$. The dash-dotted curve corresponding to the label on the right is the charge carried by the excitation $\Delta C$. (b) The corresponding quasiparticle excitation spectrum. The dashed horizontal line decribes the chemical potential $\mu=0$. The other parameters for both figures are set as: $\phi =\frac{\pi}{2}$, $t_{2\uparrow} = 0.2t_{1\uparrow}$.   }
\label{excitation}
\end{figure}

 The Haldane-Hubbard FKM differs from the normal Haldane model in the existence of a solitonic excitation in the present model generated by flipping a $Z_2$ variable $D_i$ from the ground state configuration. In Fig. \ref{excitation}, we plot the corresponding excitation energy, the charge and the spin carried by the excitation as functions of interaction $U$ when we flip the $Z_2$ number from $-\frac{1}{2}$ to $\frac{1}{2}$ on one site. We set $\phi =\frac{\pi}{2}$ and $t_{2\uparrow} = 0.2t_{1\uparrow}$ in our calculation.


From Fig. \ref{excitation}(a) we can see that the solitonic excitation undergoes a transition at $|U|\simeq 3 t_{1,\uparrow}$. When $|U|\leq 3 t_{1,\uparrow}$, the solitonic excitation energy increases parabolically as $|U|$ increases. The charge carried by the excitation is equal to $1$ and the spin of the excitation equals $-\frac{1}{2}$, corresponding to trapping an additional spin down fermion on that site. When $|U|\geq 3 t_{1,\uparrow}$, the excitation energy starts to decrease as interaction gets stronger. In this region, the charge excitation jumps to $0$ or $2$ and the spin excitation jumps to $-1$ or $(0)$, depending on the sign of $U$, indicating that a spin-up hole(electron) is also trapped on that site to lower the energy when $|U|$ is large enough. 

 It is interesting to note from Fig. \ref{excitation}(b) that the quasi-particle gap first closes and then reopens with the increase of $U$, indicating a topological phase transition. The topological phase transition is accompanied by the appearance of an in-gap state after the gap reopens (topological trivial region),  the in-gap state is occupied when its energy is below the chemical potential line. The crossing point corresponds to the transition point for the topological excitation displaced in Fig. \ref{excitation}(a).

\begin{figure}[h!]
\centering
\includegraphics[width=0.7\columnwidth, height = 1.8 in]{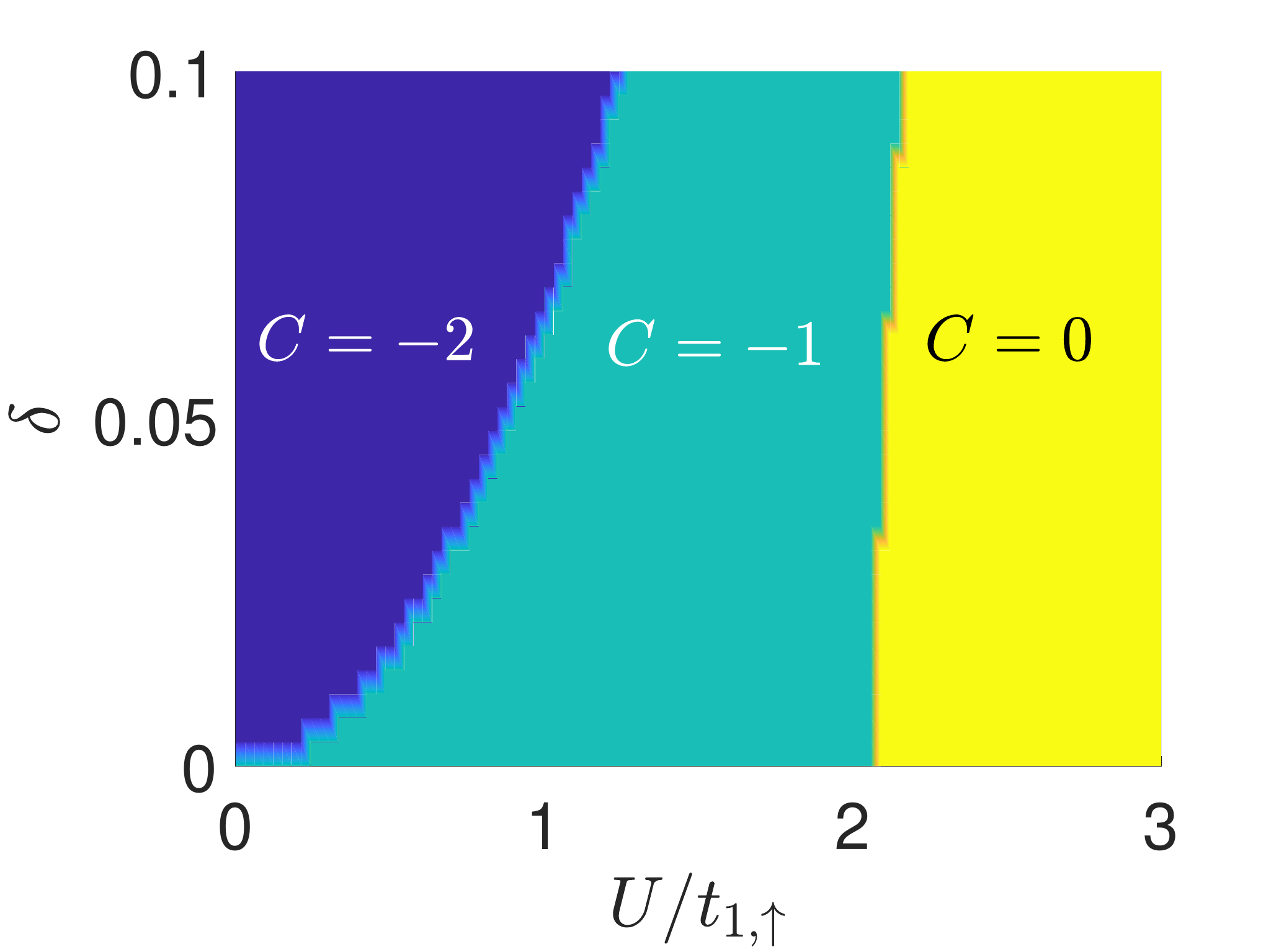}
\caption{Chern number phase diagrams as a function of $U/t_1$ and $\delta$. We set $\phi = \frac{\pi}{2}$ and $t_{2,\sigma} = 0.2 t_{1,\sigma}$. }\label{Fig:ChernPhaseDiagram}
\end{figure}

Experimentally it is extremely hard to prepare a system in the exact FKM limit. Denoting $\delta_i = \frac{t_{i,\downarrow}}{t_{i,\uparrow}} (i=1,2)$, a more realizable situation is $0<\delta_i \ll 1$. In this case, the system is not exactly solvable. However if $\delta_i$'s are small enough, one can imagine that the system deviates just a little bit from the exactly solvable limit and we can build a self-consistent mean-field theory for the system treating $\langle D_i\rangle$ as a mean-field variable. The mean-field theory becomes exact in the limit $\delta_1=\delta_2=0$. We have performed such a mean-field calculation and our result is presented in Fig. \ref{Fig:ChernPhaseDiagram} which shows how the Chern number of the system changes as we change $U/t_{1,\uparrow}$ and $\delta_1=\delta_2=\delta$ with fixed $\phi = \frac{\pi}{2}$ and $M=0$. We see that with the increase of $\delta$, the region with $C=-2$ becomes larger while the $C = -1$ region shrinks.  In the small $U$ limit, when $\delta > 0 $ the system can be viewed as two coppies of normal Haldane models with different bandwidth hence $C=-2$. With the increase of interaction, the repulsion causes slight difference of density between sublattices which makes the flatter band (spin-down band here) topological trivial. Thus the system goes into $C=-1$ phase. When increasing $U$ further, the other band will also become topological trivial due to the same reason. 

We note that the Hamiltonian (\ref{HU}) becomes an $XXZ$ spin model [\onlinecite{liu2015quantum}] in the large (positive) $U$ limit at half-filling and further reduces to the classical Ising model in the FKM limit with
\[
H_{eff} = \sum_{\langle i,j \rangle} \frac{t_{1\uparrow} ^2}{U} S_i^z S_j^z + \sum_{\langle \langle i,j \rangle \rangle} \frac{t_{2\uparrow} ^2}{U} S_i^z S_j^z
 \]
because the $\uparrow$-spins are localized. When $t_{2\uparrow} \ll t_{1\uparrow}$, the first term dominates and thus the ground state will be anti-feromagnetic. When $t_{2\uparrow} \gg t_{1\uparrow}$, the second term dominates and the system becomes frustrated. This model has been studied in Ref. [\onlinecite{houtappel1950order}] and we refer the readers to it for further details. 

\subsection{BCS-Hubbard Model with $P$-wave Pairing}
\begin{figure}[h!]
\centering
\includegraphics[width=0.6\columnwidth]{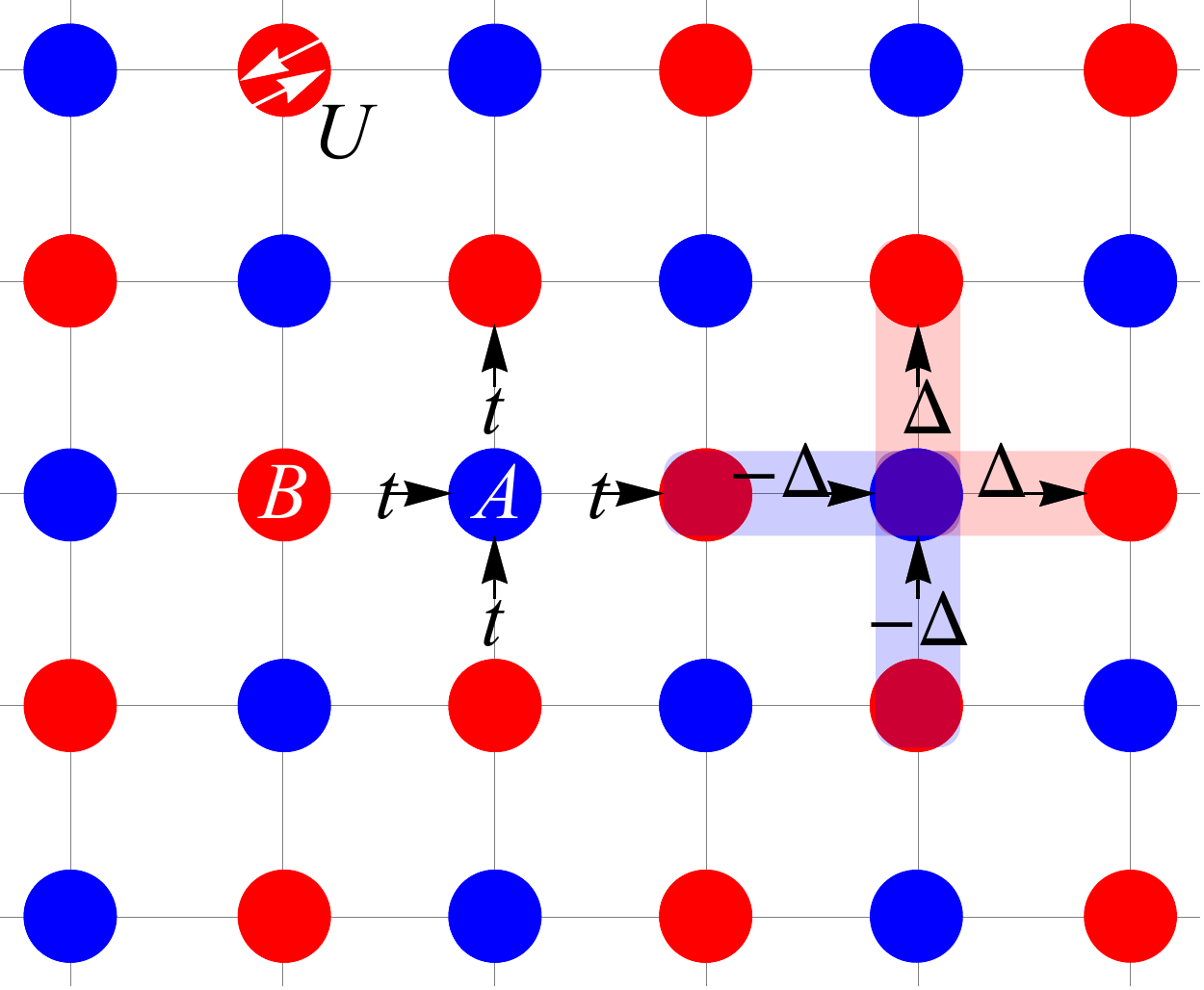}
\caption{BCS-Hubbard model on square lattice with the blue dots representing $A$ sub-lattice and red dots representing $B$ sub-lattice  [\onlinecite{chen2018exactly}]. The hopping term $t$ is uniform along all the nearest neighbor bond, i.e. $t_{A\rightarrow B} = t_{B\rightarrow A}$, while the pairing potential has a staggered form  $\Delta_{A\rightarrow B} = -\Delta_{B\rightarrow A}$. $U$ represents the Hubbard on-site interaction.}\label{Fig:BCSModel}
\end{figure}
 We next consider the BCS-Hubbard model on a square/cubic lattice given by $H=H_0+H_{int}$ stated in Ref. [\onlinecite{chen2018exactly}], with
\begin{equation}\label{h}
\begin{split}
H_0 = & \sum_{\langle i,j\rangle,\sigma}\left(t_{ij}c^{\dagger}_{i\sigma}c_{j\sigma}+h.c. +\Delta_{ij} c^{\dagger}_{i\sigma}c^{\dagger}_{j\sigma}+h.c.\right) \\
H_{int} = & U\sum_{l}(n_{l\uparrow}-{1\over2})(n_{l\downarrow}-{1\over2})
\end{split}
\end{equation}
 where $\langle i,j\rangle$ describes nearest neighbor sites with $i\in A, j\in B$ being lattices sites belonging to different sublattices on a cubic or square lattice. $t_{ij}=t_{ji}=t$ and $\Delta_{ij}=-\Delta_{ji}=\Delta$ are hopping matrix and BCS-pairing term between sites $i$ and $j$, respectively. The last term describes on-site Hubbard interaction $U$ where $l\in A,B$, i.e. all lattices sites where $n_{l\sigma}=c^{\dagger}_{l\sigma}c_{l\sigma}$. The BCS-pairing term describes a p-wave superconductor with equal spin pairing (ESP). We shall consider a pairing term $\Delta_{ij}=\Delta$ which is positive when $i\in A, j\in B$ (see Fig.  \ref{Fig:BCSModel}). Through the construction of Majorana fermions and a staggered $\mathrm{SO}(4)$ rotation corresponding to {\textbf{Case b}} of Sec. \ref{Sec:MaFKM}, the Hamiltonian\ (\ref{h}) can be transformed to a simple form

\begin{eqnarray}
\label{hma}
H_0 & \rightarrow & 4i\tilde{t}\sum_{\langle i,j\rangle,\sigma}\left(-\gamma_{i\sigma}\gamma_{j\sigma}+\delta\eta_{i\sigma}\eta_{j\sigma}\right) \\ \nonumber
H_{int} & \rightarrow & U\sum_{l}(2i\eta_{l\uparrow}\gamma_{l\uparrow})(2i\eta_{l\downarrow}\gamma_{l\downarrow})
\end{eqnarray}
where $t=\tilde{t}(1+\delta)$ and $\Delta=\tilde{t}(1-\delta)$. In the limit $t=\Delta$ (or $\delta=0$), the kinetic ($H_0$) term is expressed in terms of Majorana fermions $\gamma$'s only. The Hamiltonian describes an extended MFKM.

\begin{figure}[h!]
\centering
\includegraphics[width=0.7\columnwidth, height=1.8in]{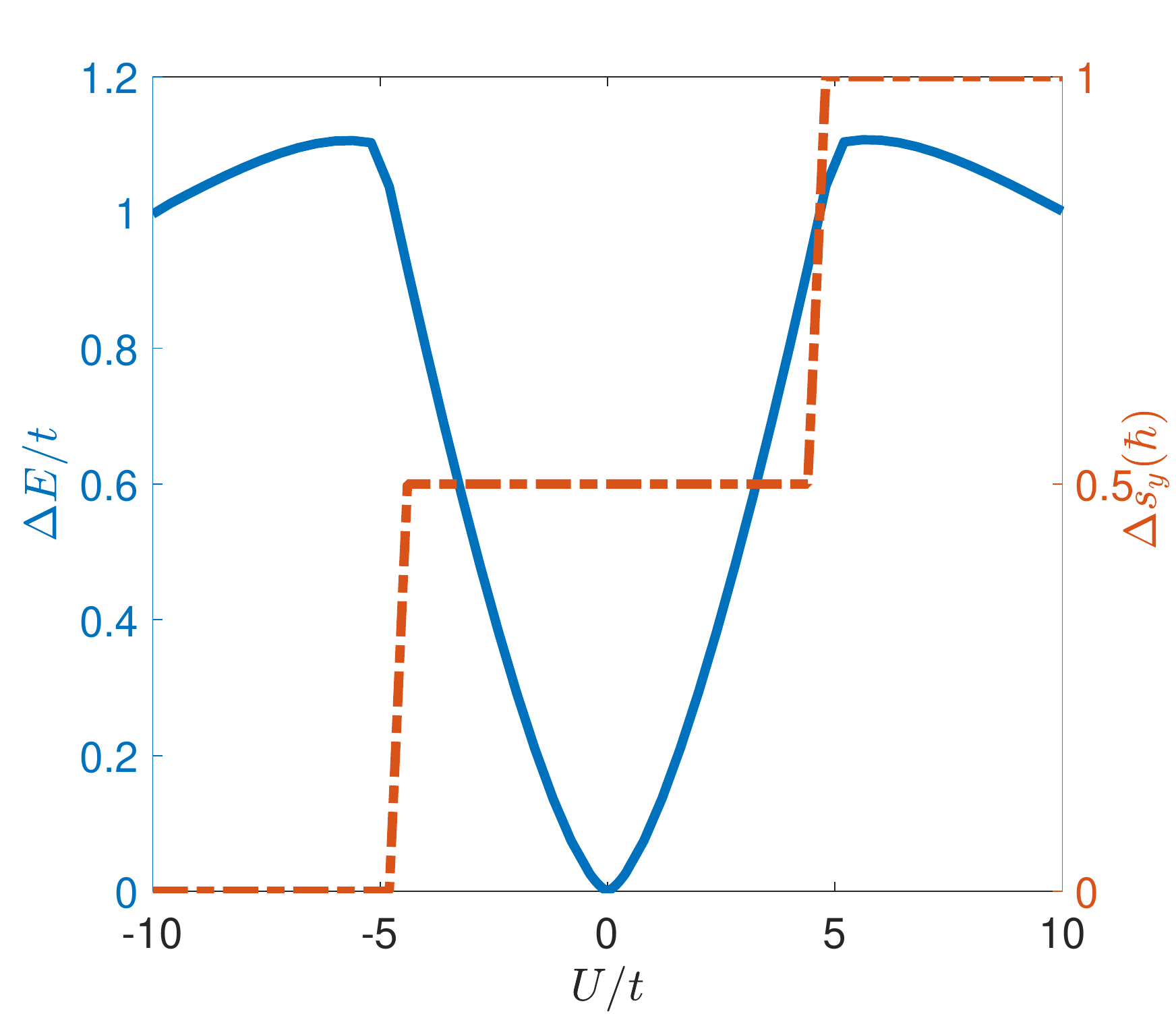}
\caption{The solitonic excitation in BCS-Hubbard model. The solid curve corresponding to the left label is the solitonic excitation energy $\Delta E$. The dash-dotted curve corresponding to the right label is the spin $\Delta m_y$ carried by the excitation. The charge carried by the excitation is always 0.}\label{Fig:MFKMExcitation}
\end{figure}

The above model exhibits interesting properties. It has spin-$1/2$ chargeless quasi-particles and solitonic excitations which may be fermions or bosons, depending on the strength of interaction. The energy and spin carried by the solitonic excitation are shown in Fig. \ref{Fig:MFKMExcitation}. The main difference between this model and the Haldane-Hubbard model described in Sec. \ref{Sec:FKMHaldane} is that here the charge excitation is always zero. This is because the quasi-particles are equal superposition of particle- and hole- excitations in the present model. Furthers details can be found in Ref. [\onlinecite{chen2018exactly}].

We note that an extension of this model that describes a topological superconductor has also been proposed recently by introducing a next nearest neighbor hopping and pairing in a honeycomb lattice [\onlinecite{ezawa2018exact}].

It is interesting to examine the effect of particle-hole transformation described in Sec. \ref{Sec:MaFKM} on this model. Consider
\begin{equation}\label{Eq:PHU2_2}
\left( \begin{matrix}
c_{\uparrow} \\
c_{\downarrow}^\dagger
\end{matrix}\right) =  \mathcal{R}_{ph} \left( \begin{matrix}
\tilde{c}_\uparrow \\
\tilde{c}_\downarrow^\dagger
\end{matrix}\right),\ \mathcal{R}_{ph}\equiv \left(\begin{matrix}
a & b \\
-b^* & a^*
\end{matrix} \right)
\end{equation}
where $a=1, b=0$ on A-sublattice and $a=0, b=1$ on B-sublattice. In terms of $\tilde{c}$- fermions, the model becomes
\begin{equation}\label{h1}
\begin{split}
H  \rightarrow & \sum_{\langle i,j\rangle}\left(t_{ij}(\tilde{c}^{\dagger}_{i\uparrow}\tilde{c}^{\dagger}_{j\downarrow}-
\tilde{c}^{\dagger}_{i\downarrow}\tilde{c}^{\dagger}_{j\uparrow})+h.c. \right. \\ \nonumber
&  \left.+\Delta_{ij} (\tilde{c}^{\dagger}_{i\uparrow}\tilde{c}_{j\downarrow}-\tilde{c}^{\dagger}_{i\downarrow}\tilde{c}_{j\uparrow})+h.c.\right) \\  \nonumber
 &  + U\sum_{l}(\tilde{n}_{l\uparrow}-{1\over2})(\tilde{n}_{l\downarrow}-{1\over2}).
 \end{split}
\end{equation}
Notice that the hopping term becomes an extend-$s$-wave BCS pairing term and the pairing term becomes a vector (spin triplet) hopping term under this transformation. This is not surprising since our model breaks rotation symmetry and $s-$ and $p-$ wave pairings are in general mixed in our model. Notice also that under the particle-hole transformation, the charge carried by the $c$- and $\tilde{c}$- fermions are the same on $A$-sublattice, but are opposite on $B$-sublattice. For more general particle-hole transformations, the charge carried by the $c$- fermions may be a fraction of the charge carried by the $\tilde{c}$- fermions and vice versa as we shall see in the next subsection.


\subsection{Anderson Lattice Model}\label{Sec:AndersonLattice}
In this subsection, we consider two related examples of BCS-Anderson lattice models. The models are extended MFKM's as discussed in Sec. \ref{Sec:AndersonModel} . The first model we consider has Hamiltonian
\begin{equation}\label{Eq:Anderson}
\begin{split}
H = &t\sum_{\langle i,j \rangle,\sigma'\sigma}  (c^\dagger_{i,\sigma} c_{j,\sigma}  + h.c.)+\sum_{i,\sigma} \epsilon_{\sigma} (c^\dagger_{i,\sigma}c_{i,\sigma}) \\& + U\sum_{i} (f_{i,\uparrow}^\dagger f_{i,\uparrow}-\frac{1}{2})(f_{i,\downarrow}^\dagger f_{i,\downarrow}-\frac{1}{2})\\&+ H_{c}
\end{split}
\end{equation}
where $\sigma =\uparrow,\downarrow$ are spin variables . $c(c^\dagger)$ is the annihilation (creation) operator for $c$-fermions while $f(f^\dagger)$ is the annihilation (creation) operator for localized $f$-fermion. $\langle i,j \rangle$ denotes nearest neighbors on the lattice. $\epsilon_\sigma=\mu+\sigma B$ represents a spin-dependent on-site energy for $c$-fermions. The on-site energy term is set to be zero for the $f$-fermions. There is an on-site Hubbard interaction $U$ between $f$-fermions.

  $H_c$ describes an onsite coupling between the $c$- and $f$-fermions with
  \[ H_c=i\frac{V}{2}\sum_{i,\sigma}(c_{i,\sigma}+c_{i,\sigma}^\dagger)(f_{i,\sigma}+f_{i,\sigma}^\dagger) \]
  in our first model.  In the basis $\Psi_c = \{c_{\uparrow},c_{\downarrow},c_{\uparrow}^\dagger ,c_{\downarrow}^\dagger\}^T$ and
$\Psi_f = \{f_{\uparrow},f_{\downarrow},f_{\uparrow}^\dagger ,f_{\downarrow}^\dagger\}^T$, this term is represented as $H_c=\sum_i\Psi_{ci}^\dagger h_c \Psi_{fi}$ where
\begin{equation}\label{Eq:hc}
h_{c} =   i\frac{V}{2}\left(\begin{matrix}
1 & 0 & 1 &0\\
0 & 1 & 0 &1\\
1 &0 & 1 &0\\
0 & 1 & 0 &1
\end{matrix}\right).
\end{equation}
In this case, the Majorana fermions $\xi_{i\sigma}=i(f_{i,\sigma}-f^\dagger_{i,\sigma})$ does not enter $H_c$ and $D_i = i\xi_{i\uparrow}\xi_{i\downarrow}$ is a constant of motion in the Hubbard interaction term. The model reduces to an extended MFKM. The ground state of the system is determined by the configuration of $D_i$'s that minimizes the total energy. We have performed the calculation numerically for a one-dimensional lattice with periodic boundary condition and find that the ground state configuration is fully determined by the relation between chemical potential $\mu$ and Zeeman field $B$. When $\mu>B$ the system favors uniform configuration $D_i=D \forall i$.  When $B > \mu$ the system favors staggered configuration $D_i=(-1)^i D$. The uniform and staggered configurations are degenerate when $\mu = B$.  The relation between $B>(<)\mu$ can be understood by making a transformation for the system
\begin{equation}\label{Eq:Staggered_Transform}
\begin{split}
&\tilde{c}_{i\uparrow} = c_{i\uparrow}, \tilde{c}_{i\downarrow} = (-1)^{i}c_{i\downarrow}^\dagger, \tilde{f}_{i\uparrow} =f_{i\uparrow}, \tilde{f}_{i\downarrow} = (-1)^{i}f^\dagger_{i\downarrow}, \\
&\tilde{U} = -U, \tilde{\mu} = B, \tilde{B} = \mu.
\end{split}
\end{equation}
It is easy to show that the Hamiltonian (\ref{Eq:Anderson}) is invariant under the transformation. In particular, the chemical potential becomes the magnet field term in the transformed basis and  vice versa, i.e., there exists a one-to-one mapping between the ground states at $B>\mu$ and at $\tilde{\mu}>\tilde{B}$. In particular, when $\tilde{D}_i = i\tilde{\xi}_\uparrow \tilde{\xi}_\downarrow$ is uniform, then $D_{i}=(-1)^{i}\tilde{D}_i = (-1)^{i}D$ is staggered.

\begin{figure}[h!]
\centering
\mbox{
\subfigure{
\includegraphics[width=0.5\columnwidth, height=1.5in]{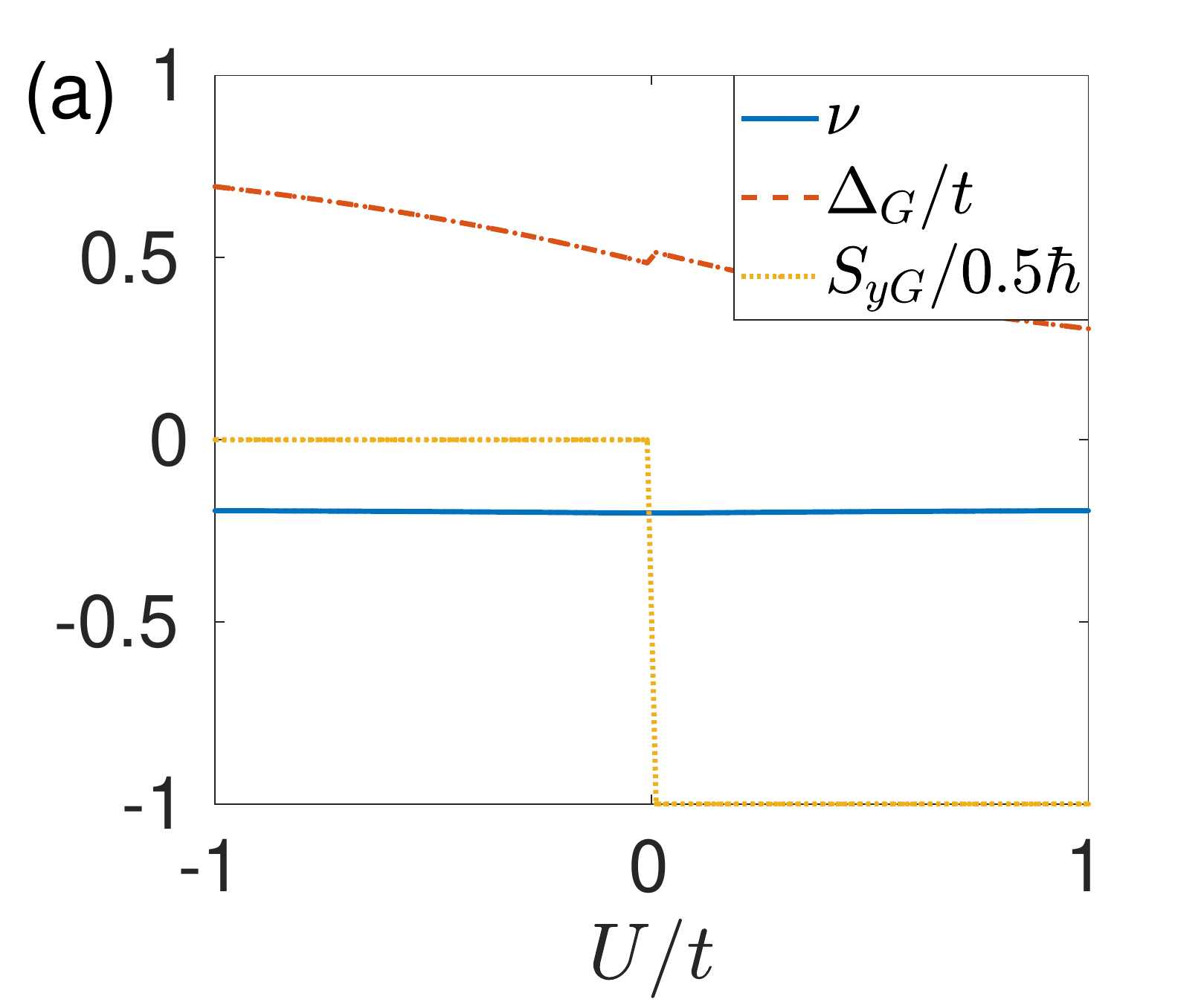}}
\subfigure{
\includegraphics[width=0.5\columnwidth, height=1.5in]{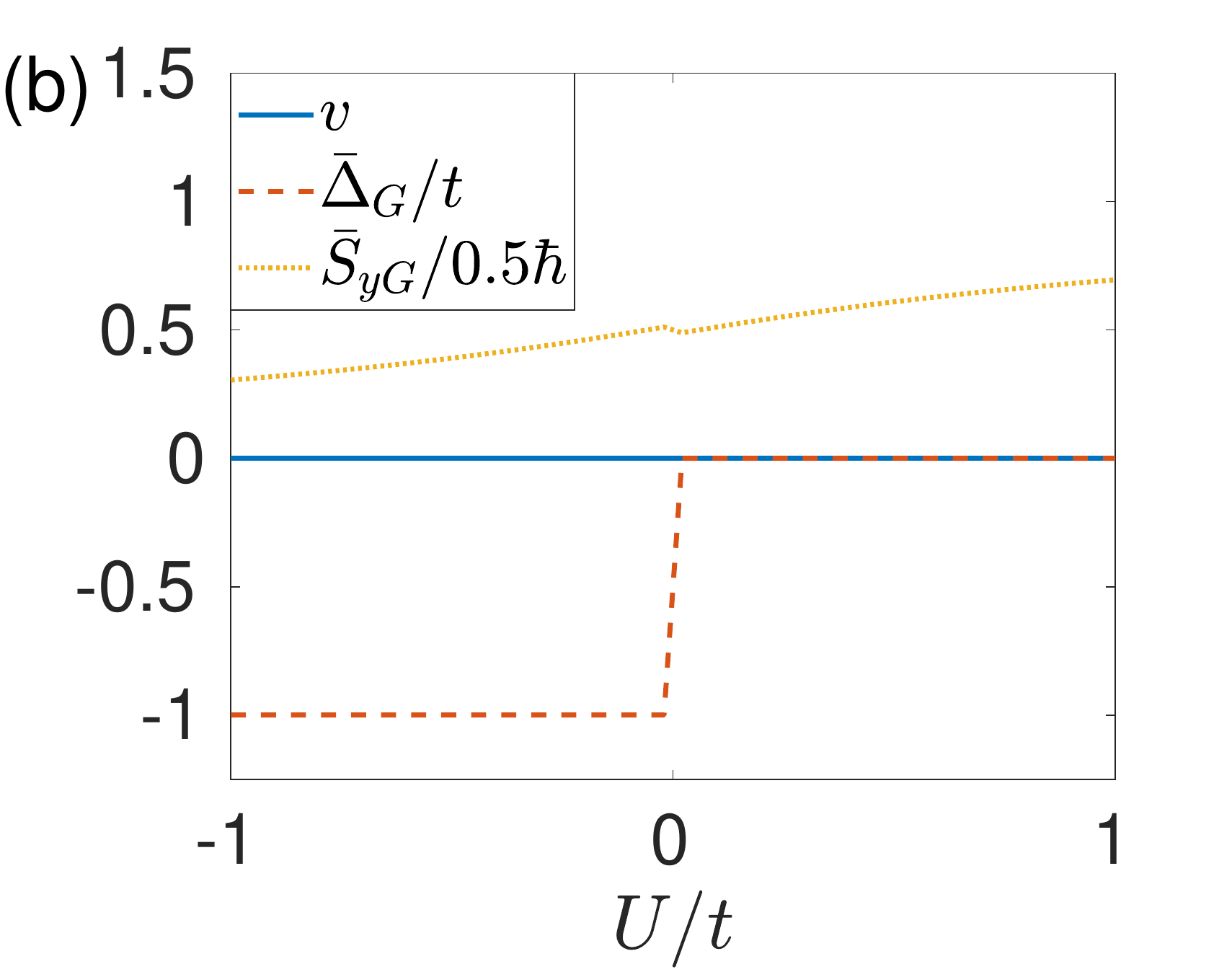}}}
\caption{(a) The ground state properties as a function of interaction strength $U$ for $\mu=0.3t$ and $B=0$, ($\mu > B$, the uniform  $D_i$ case). The solid curve represents the total electron density of the system $\nu$, the dashed curve represents the pairing correlation $\Delta_G/t$ while the dotted curve represents the magnetization along $y$ direction $S_{yG}/0.5\hbar$. (b) The ground state properties as a function of interaction strength $U$ for $\mu =0$ and $B = 0.3 t$ ($\mu<B$,  the staggered  $D_i$ case). The solid curve represents the total electron density of the system $\nu$. The dashed curve represents the staggered pairing correlation $\bar{\Delta}_G/t$ while the dotted curve represents the staggered magnetization along $y$ direction $\bar{S}_{yG}/0.5\hbar$. The other parameters  are set as $V=0.6t$, $N=50$. }\label{Fig:AndersonGround}
\end{figure}

 To compare the difference between the two cases, we compute the ground states uniform and/or staggered magnetization along $y$-direction and on-site $s$-wave pairing amplitude, defined as follows:

\begin{subequations}
\begin{eqnarray}
\Delta_G =&\frac{1}{N}\sum_i \langle G|\hat{\Delta}_i  |G\rangle \\
S_{yG} =&\frac{1}{N}\sum_i \langle G| \hat{S}_{yi}  |G\rangle \\
\bar{\Delta}_{G} =&\frac{1}{N}\sum_i  (-1)^i\langle G| \hat{\Delta}_i   |G\rangle \\
\bar{S}_{yG} =&\frac{1}{N}\sum_i (-1)^i\langle G| \hat{S}_{yi}  |G\rangle \\
\nu = &\frac{1}{N}\sum_{i} \langle G | \hat{n}_{if}+\hat{n}_{ic}-1 | G\rangle
\end{eqnarray}
\end{subequations}
 where $|G\rangle$ is the ground state wave function, $N$ is the number of total sites. $\hat{\Delta}_i=\hat{\Delta}_{fi}+\hat{\Delta}_{ci}$ where $\hat{\Delta}_{c(f)i}=i(c(f)_{i\uparrow}c(f)_{i\downarrow}-c(f)_{i\downarrow}c(f)_{i\uparrow})$ is the $s$-wave pairing operator between $c(f)$ electrons on-site $i$.  $\hat{S}_{yi}=\hat{S}_{yi}^{(c)}+\hat{S}_{yi}^{(f)}$ where $\hat{S}_{yi}^{c(f)} = i c^\dagger(f)_{i\uparrow}c(f)_{i\downarrow} +h.c.$ is the magnetization at site $i$ carried by $c(f)$ electrons along $y$ direction. (To avoid confusion we note here that the external magnetic field $B$ is applied along $z$-direction. The Majorana fermion hybridization term breaks spin-rotation symmetry and induces magnetization along $y$-direction.) The ground state properties corresponding to the two configurations(uniform and staggered) are shown in Fig. \ref{Fig:AndersonGround}(a) and Fig. \ref{Fig:AndersonGround}(b), respectively.

 For both Fig. \ref{Fig:AndersonGround}(a) and Fig. \ref{Fig:AndersonGround}(b), we first notice that there exists a first order phase transition at $U=0$, making $U=0$ a critical point. For Fig. \ref{Fig:AndersonGround}(a) ($\mu>B$),  the ground state magnetization along $y$-direction changes from zero to ferromagnetic as the interaction changes from negative to positive, the uniform superconducting order parameter remaining finite throughout the phase transition. We note that although $H_c$ drives a $p$-wave superconducting order, the $s$-wave pairing amplitude is nonzero because the Majorana fermion hybridization term breaks rotation symmetry. In Fig. \ref{Fig:AndersonGround} (b) we show the case for $\mu<B $. In this case, the ground state $s$-wave superconducting order changes from staggered non-zero to zero as the interaction changes from negative to positive, with a staggered magnetization  (along $y$-direction) remaining throughout the phase transition. This behavior can be understood from the transformation\ (\ref{Eq:Staggered_Transform}).  We can see that the staggered pairing correlation $\Delta_G/t$ becomes the magnetization $S_{yG}/(0.5\hbar)$ if $U\rightarrow -U$ while staggered magnetization $\bar{S}_{yG}/0.5\hbar$ becomes the (uniform) pairing correlation. We note also that in the first case (Fig. \ref{Fig:AndersonGround}(a)) the electron density is nonzero (away from half-filling) because $\nu\neq0$.

Next we consider our second model.  The model has the same form as Eq.\ (\ref{Eq:Anderson}) except that $H_c=\Psi_c^\dagger \tilde{h}_c \Psi_f$ with
\begin{equation}
\tilde{h}_{c} =   i\frac{V}{2}\left(\begin{matrix}
1 & i & 1 &-i\\
-i & 1 & i &1\\
1 &i & 1 &-i\\
-i & 1 & i &1
\end{matrix}\right)
\end{equation}

It is straight forward to show that $\tilde{h}_c$ can be generated from $h_c$ through a transformation $\mathcal{P}_f=\mathcal{P}_{ph}\otimes\mathcal{P}_s$ defined in Eq. (\ref{Eq:P_ph}) and Eq. (\ref{Eq:P_s})with $\mathcal{P}_{ph} = \frac{1}{\sqrt{2}}(\rho^0\sigma^0+i\rho^x\sigma^y)$ and $\mathcal{P}_s = \mathbb{I}$, corresponding to a particle-hole transformation for the $f$-fermions
\begin{equation}
\label{Eq:phanderson}
\left( \begin{matrix}
\tilde{f}_{\uparrow} \\
\tilde{f}_{\downarrow}^\dagger
\end{matrix}\right) = {1\over\sqrt{2}}\left(\begin{matrix}
1 & i \\
i & 1
\end{matrix} \right)
\left( \begin{matrix}
f_\uparrow \\
f_\downarrow^\dagger
\end{matrix}\right).
\end{equation}
  The Hamiltonian is thus the same as Eq.\ (\ref{Eq:Anderson}) except $f(f^\dagger)_{\sigma}\rightarrow\tilde{f}(\tilde{f}^\dagger)_{\sigma}$ ($\sigma=\uparrow\downarrow$).

\begin{figure}[h!]
\centering
\mbox{
\subfigure{
\includegraphics[width=0.5\columnwidth, height=1.5in]{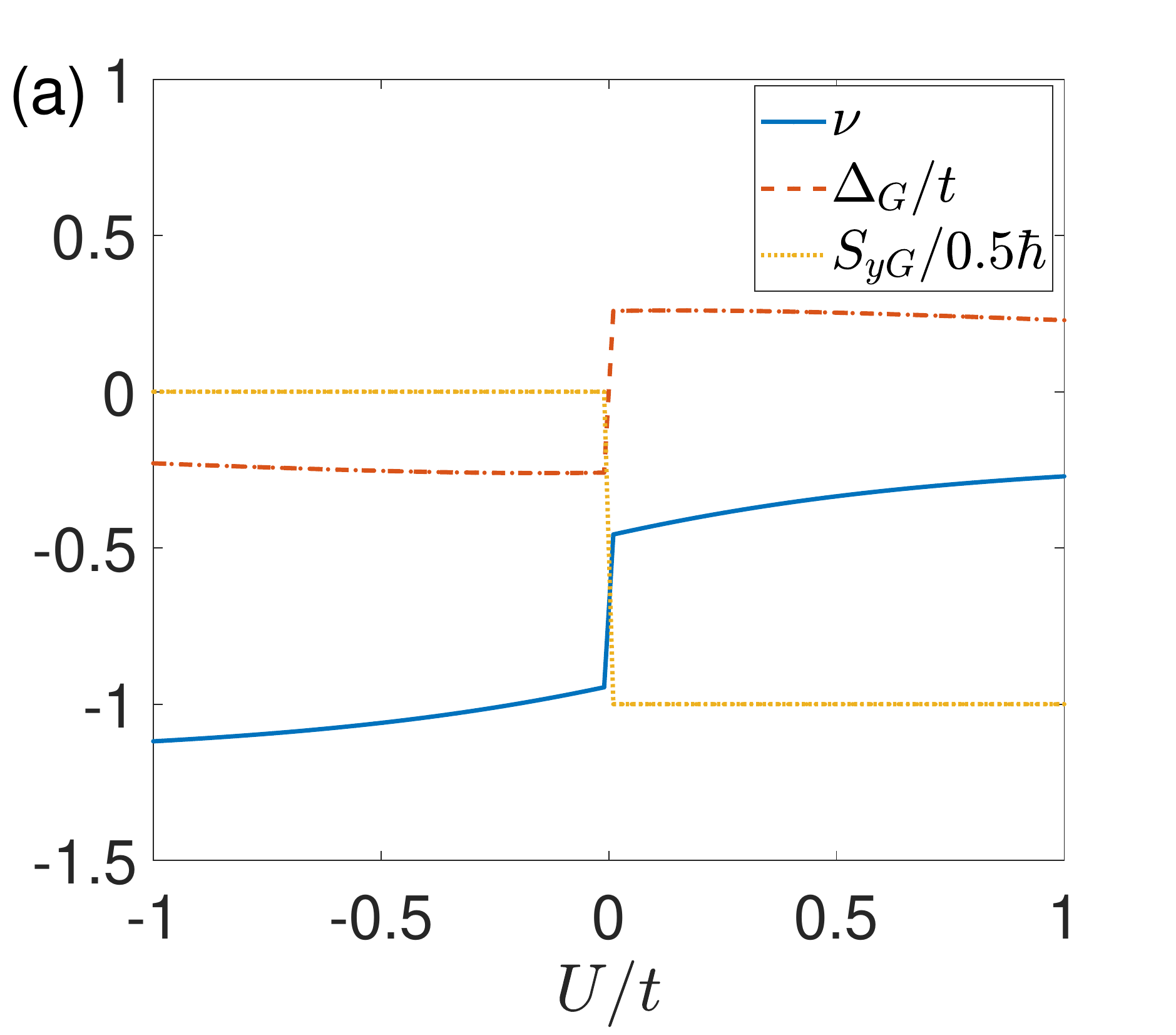}}
\subfigure{
\includegraphics[width=0.5\columnwidth, height=1.5in]{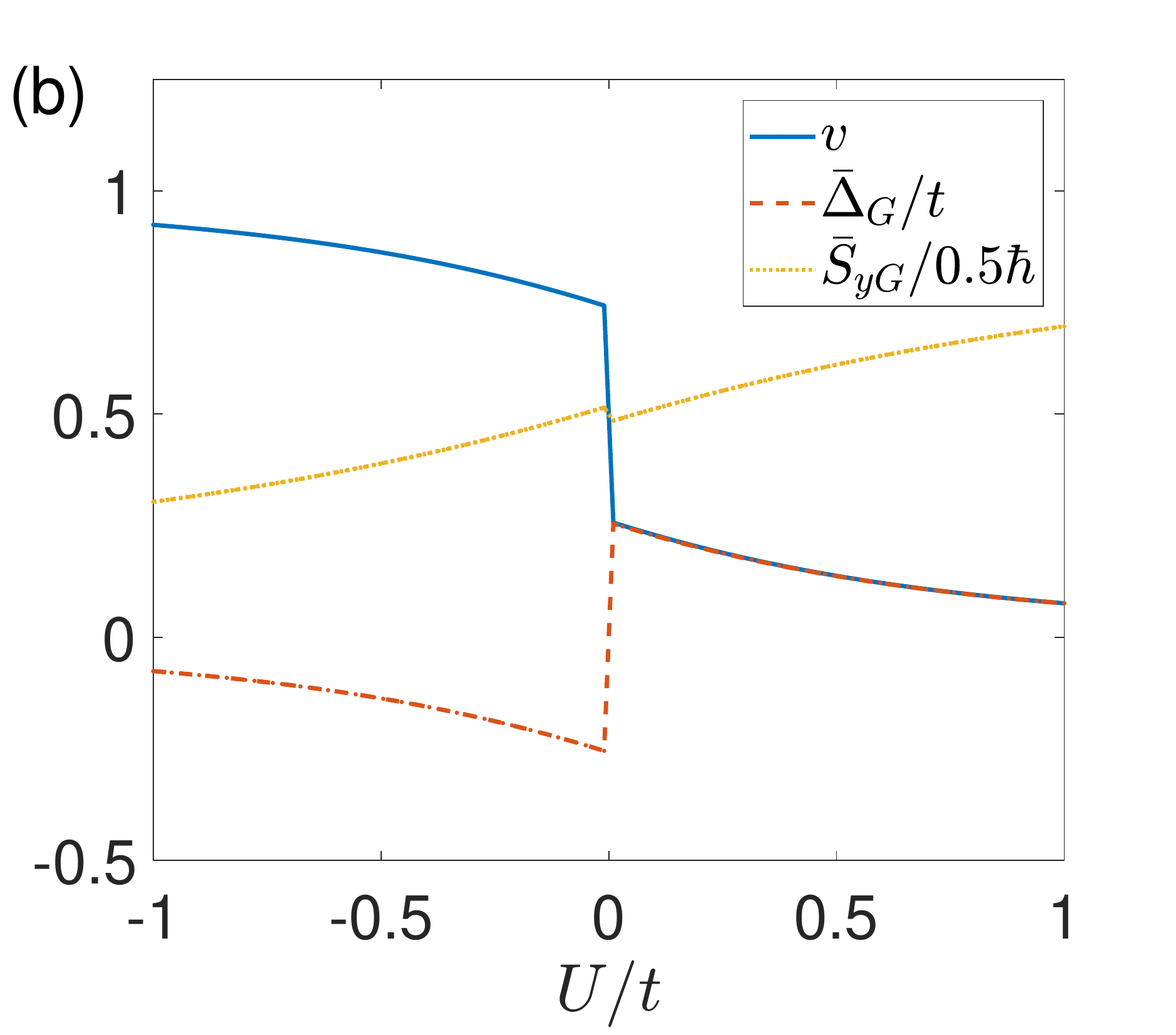}}}
\caption{(a) The ground state properties as a function of interaction for the uniform  $D_i$ case ($\mu > B$, here $B =0$ and $\mu = 0.3 t$ ). The solid curve represents the filling factor of the system $\nu$. The dashed curve represents the pairing correlation $\Delta_G/t$ while the dotted curve represents the magnetization along $y$ direction $S_{yG}/0.5\hbar$. (b) The ground state properties as a function of interaction for the  staggered $D_i$ case ($\mu < B$, here $\mu =0$ and $B = 0.3 t$ ). The solid curve represents the filling factor of the system $\nu$. The dashed curve represents the staggered pairing correlation $\bar{\Delta}_G/t$ while the dotted curve represents the staggered magnetization along $y$ direction $\bar{S}_{yG}/0.5\hbar$. The other relavant parameters for both figures are set as: $V=0.6t$, $N=50$. }\label{Fig:AndersonGround2}
\end{figure}

  Although the Hamiltonian has the same form  as the previous case in the rotated basis $f\rightarrow\tilde{f}$, the physical properties of the ground state are different because the charge carried by $\tilde{f}$-fermions are not the same as the charge carried by the $f$-fermions. As is shown in Fig. \ref{Fig:AndersonGround2}, we see that the total electron density of the system $\nu=\sum_{\sigma}(\langle f^\dagger_{i\sigma}f_{i\sigma}\rangle + \langle c^\dagger_{i\sigma}c_{i\sigma}\rangle)$ changes as interaction strength $U$ changes, although $\tilde{\nu}=\sum_{\sigma}(\langle \tilde{f}^\dagger_{i\sigma}\tilde{f}_{i\sigma}\rangle + \langle c^\dagger_{i\sigma}c_{i\sigma}\rangle)$ remains unchanged as a function of $U$. The differences between (a)($\mu>B$) and (b) ($\mu<B$) can again be understood from the transformation\ (\ref{Eq:Staggered_Transform}).

\section{Conclusion}
In summary, we explore in this paper generalizations of the Falicov-Kimball model. We consider generalized Falicov-Kimball model in three types of systems:(i) normal Hubbard Hamiltonian with spin-dependent hopping; (ii) BCS-Hubbard Hamiltonian and (iii) Anderson lattice model. We explore the criteria when these three types of models can be transformed into effective Falicov-Kimball models (or extended Majorana Falicov-Kimball model). Besides ordinary electrons, we point out that the extended Majorana Falicov-Kimball model can describe spin-$1/2$ systems. Corresponding to the three general classes, we study three specific models in detail:(i) The spin-dependent Haldane Hubbard model, in which an interaction-driven topological phase transition is found. With different interaction strength, the solitonic excitations in the model carry different spins and charges, and can be fermion- or boson- like;  (ii) a $p$-wave BCS superconductor with staggered pairing potential. When the pairing potential $\Delta$ equals the hopping amplitude $t$, the system becomes exactly solvable and becomes an extended Majorana Falicov-Kimball model; (iii) an Anderson lattice model with a Majorana fermion coupling between the localized fermions and the itinerant electrons.

We emphasize here that our general criteria for finding exactly-solvable, extended Falicov-Kimball model provide guidance to find exactly solvable interacting fermions (or spin) models.  A common feature of these models is that they exhibit both quasi-particle and solitonic excitations. We note that a very recent work discusses an exactly solvable two-dimensional topological superconductors with Hubbard interaction [\onlinecite{ezawa2018exact}], the exact solution to Haldane-Hubbard-BCS-model  [\onlinecite{miao2019exact}] and the generalized Kitaev model to three dimension [\onlinecite{miao2018exact}] are examples of the extended Majorana Falicov-Kimball model we discussed.

\section{Acknowledgement}
X.H. Li, Z. Chen and T.K. Ng  thank the support of HKRGC through HKUST3/CRF/13G and C6026-16W.

\bibliography{exact}
\end{document}